# Real Signals in the Microcosm

S.G. Basiladze

Moscow State University, Institute of Nuclear Physics, Moscow 119899, Russia

**ABSTRACT**

It is known that the transition from quantum to classical physics occurs when the action constant *h* decreases to an infinitesimal value. As shown in the article, for signals this statement is applicable to their classical model only and it is far from all real signals existing in nature and having a threshold on action more than *h*. Such threshold is the threshold of perception that exists for any natural macro and micro signal receivers, including atoms and particles. It was revealed that above the boundary frequency of the classical thermal noise $\nu_H$, found by Planck, the properties of real signals correspond to quantum constraints. It is asserted that in the frequency range above $\nu_H$ inertness appears and dominates, i.e. the integration of the signals. As a result of it a transition from the perception threshold on energy to the perception threshold on action happens. Besides, instead of disappearing thermal noise on energy, an indestructible sub-threshold noise on action appears that is associated with the uncertainty relation. The concept of "algebraic" – not integer logic is introduced for real sub-threshold signals. It is demonstrated that in combination of sub-threshold random states of several signals, their integration leads to a joint probability, i.e. to superposition states of QM.

## 1. Introduction

Signals reflect of the processes occurring in nature and phenomena that are perceived by measuring instruments. Physics, like any natural science, cannot manage without the concept of a signal; it receives with signal help the knowledge about what surrounds us. The latter is usually got via the following chain: the carrier of the knowledge are the measurement data represented by the minimized code, i.e. by information [1÷4]; the information carrier are those or other <u>*states*</u> of the signals registered in <u>*events*</u> of measurement, the carrier of the signals themselves is energy.

> *Note*. Here and below in underlined italics are given the <u>*terms*</u> used in this work. Further, if it is necessary to indicate that the *term* is used, it is written in italics.

"Measuring tools" in the technology of knowledge are not created by a man; he only uses certain phenomena already existing in nature. The world around us is saturated with signals, including astro- and micro-scales. The most obvious example using the concept of a signal is the relativity theory, based on the assertion of the finiteness of the signal propagation speed. In the microcosm, each atom or particle can be a source or "receiver" some portions of energy and, by virtue of this, actually carry out the signal exchange.

*1.1. The limitation of the classical signal theory*

The existing theory of signals is constructed, like classical physics, on the basis of the concepts of <u>c</u>lassical <u>m</u>athematical <u>a</u>nalysis (CMA), for which 4 <u>c</u>lassical <u>p</u>roperties (CP) are characteristic.

**CP1**. In CMA, each point is by definition distinguishable; otherwise it is meaningless to talk about individual points.

**CP2**. Each point of the function y=f(x) can be moved move along 'y' (at x=const) by any distance up to infinity.

**CP3**. The same point can be moved infinitely close to the 'x' axis. In both cases, each shift produces a different function than the original one. The initial and "shifted" functions can be used as two distinguishable *states*.

**CP4**. For the next measurement *event*, other points that are infinitesimally close to the first one can be used; i.e. at a distance x+δx in the projection onto the x-axis, where δx→0.

Thus, on the finite intervals Δx and Δy the classical function y=f(x) has a threefold infinite number of degrees of freedom [2], i.e. threefold infinite information capacity, which is absolutely unreal.

Since the classical analog signal is represented by a continuous mathematical function y=f(x), it inherits from it the properties of CP2÷CP4, i.e. three times the infinite information capacity. If the property KC1 in CMA can simply be *declared*, then in a signal the recognition of each point must be *justified physically*.

As for CP4, it is usually assumed that according to the sampling theorem [6], in the harmonic band from 0 to $g_H$, an infinitesimal step in 'x' may be discrete: dx→δx. This means that in the sampling function only one point from its infinite set, carries an information – this is peak point (Fig. 1c), and all others points are "*empty*" (their position is predictable).

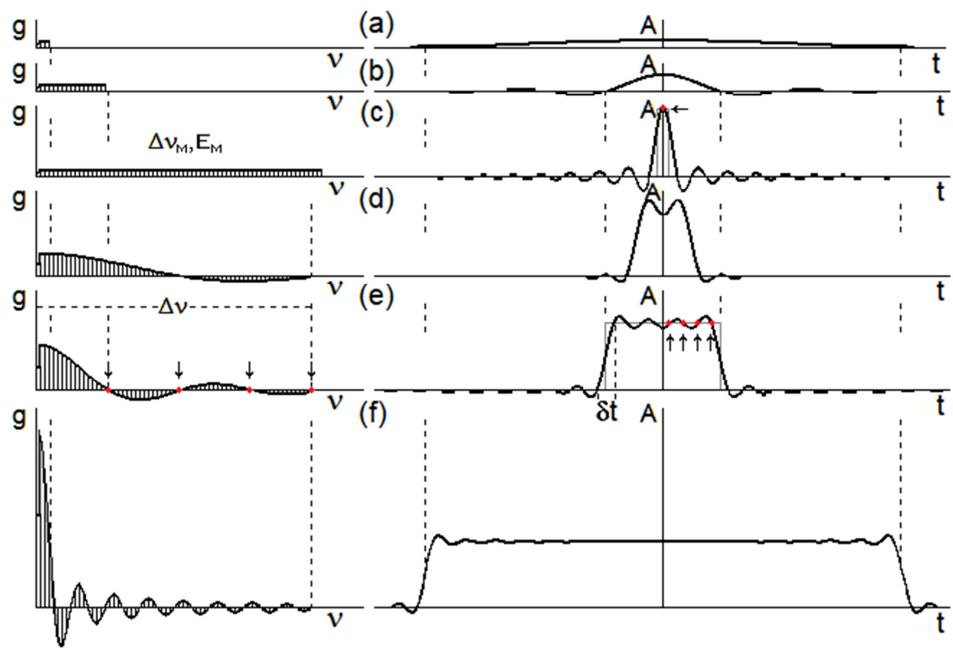

Fig. 1. (a)÷(c) An example of the growth of signal energy; signal represented by the sampling function with constant spectral energy density. (d)÷(f) The signal is decomposing into sample functions at constant energy density of the signal envelope.

The sampling functions (a)÷(c) have a flat (cutoff) spectrum; a consequence of the absence of highest harmonics are the Gibbs oscillations before (!) and after the peak of the function. As an element of the signal decomposition (d)÷(f), the sampling function has only one – peak point (c) carrying information. The number of reference points in the spectrum and in the envelope of the signal is the same – 4, this is indicated by the arrows in the fragments (c), (e). The narrower the spectrum of the mapping (c)→(a), the worse the definition of the prototype (rectangular pulse). The fragment (e) shows also the relation between the spectral width Δν and the signal switching time δt; these quantities are constant on the fragments (d)÷(f).

**CP5**. However, the sampling theorem contradicts the fact that physically, the dynamic signal spectrum (x≡t, 1/t=ν) cannot consist only of harmonics from 0 to $ν_H$, it must be infinite (more precisely, infinitely decreasing after $ν_H$). In other words, the sampling function is not physically realizable. In the absence of higher harmonics (Fig. 1b÷c) the signal, that is decomposing onto the sample functions (Fig. 1e), gets "wings" on both sides - infinite Gibbs oscillations [7] containing up to 10% of its energy. This violates the Wiener causality condition [8], since such a signal can be detected by a sensitive receiver long before its actual appearance.

**CP6**. Unfortunately, a classical signal with an infinite decreasing spectrum, satisfying the conditions of finiteness of energy and causality, acquires the property of an infinite (!) rapidity of switching (jumps) from one state to another.

The revealed set of contradictory properties is enough to doubt the applicability of classical signals in the microcosm. Nevertheless, there are 2 more points related to noise, which require clarifying.

According to Shannon [5], the property of CP3 – an infinitesimal step on 'y', cannot practically be used in signals because of the inevitable presence of classical noise, which does not allow us to work with signals whose power is less than the RMS-noise power $<P_{Ns}>$ – Fig. 2a. This means that the discreteness of states *always* appears due to noise.

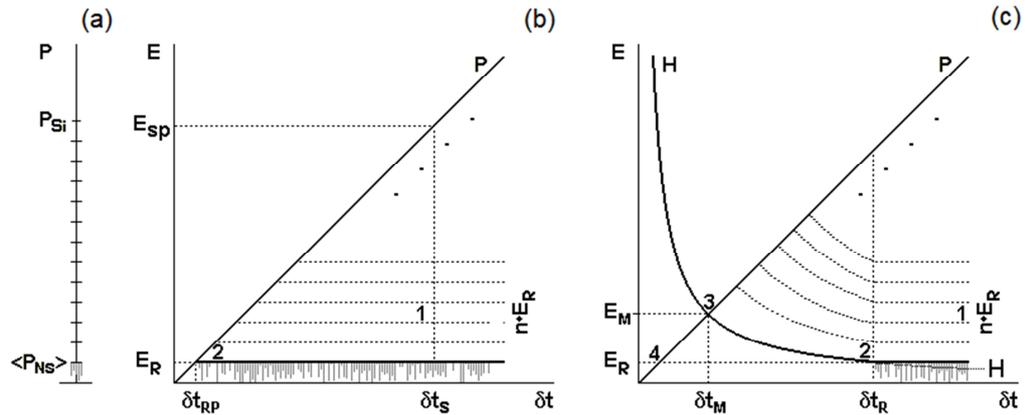

Fig. 2. The upper and lower boundaries of the signals: (a) The classical Shannon signal is limited from above by the transmitter power (density of energy) and from below – by the RMS noise value at the receiver input. (b) The zone of existence of a real *slow* signal is limited by the density of transmitted energy, and from below by the threshold of perception on energy, or by noise. (c) The left part, the region of integration of the real *fast* signal is limited from below by the perception threshold on action of the signal, classical noise is absent here.

Main notations: $\delta t_S$, $E_{SP}$ – switching time and energy of the signal; $\delta t_R$, $E_R$ – switching time and energy of the receiver; $\delta t_M$ is the minimum, physically possible time (but not 1/frequency) of the signal transmission at the power limit $P$ and the threshold on the action $H$. The frequency is determined by the point '2'.

**CP7**. The question that is quite obvious: does noise dominate always? In addition to noise, the process of threshold perception of the signal by the receiver or, more simply, the sensitivity of the latter can also serve as a limiting factor. A clear example: the gas medium, which is the carrier of acoustic signals, based on the thermal noise of chaotically colliding molecules of gas. Nevertheless, people do not hear this noise, which is several orders of magnitude smaller than the threshold of audibility; this gives the basis for the sound scale on energy (in decibels).

**CP8**. The most unpleasant point: the spectrum of classical thermal noise is flat, it cannot be infinite in any way, since this leads to infinity of energy, i.e. to the "ultraviolet catastrophe". If the spectrum is limited, then in the uppermost harmonics there is no any noise threshold of Shannon.

The contradictory properties of CP1÷8 is a consequence of abstractness (some distraction) of the classical world. The summation in it is replaced by integration, i.e. by the abstract density of discrete quantities (note that their dimensionality changes in this case). As a result of this procedure, abstract continuous functions appear with an infinite number of points not having size, which are *empty* informationally. To eliminate these contradictions, the real signals (RS), described in [2÷4], were introduced. Here we confine ourselves to a brief emphasis on their basic properties, which are important for further analysis.

*1.2. Short overview of the real signals properties*

The envelope of any analog signal (its natural representation) is described by the function y=f(x) that is supplemented by the concept of energy: $\delta E = y^2 \cdot \delta x$. In accordance with the Fourier expansion, the signal spectrum may be broadband – started with a constant (zero) component $g_0$, which is half of the first – the lowest harmonic: $g_0 = g_L/2$. Their opposite are narrow-band signals, whose spectrum does not have a constant component, is far from zero ($g_1 \gg g_L$), and its width is $\delta g \ll g_1$. Both signal representations (spectral and natural) have equal rights to exist: the theory of signals (as part of information theory) shows [7] that they contain the same number of <u>reference</u> points carrying *information* in the signal (Fig. 1e).

In dynamics ($x \equiv t$) $y^2$ is power *P*. The envelope of the broadband harmonics package has an area, the square of the area is an action *H* of the signal [1], i.e. the area is a <u>radical</u> of the action $\pm\sqrt{H}$ and it may be negative (like any integral). Narrowband signals have no action.

According to [2÷4], the transition from classical signals to real ones includes the following.

**RP1**. Transition from a one-dimensional diagram of the signal energy density and noise (Fig. 2a) to a more physical two-dimensional diagram – the rapidity and energy of the signal switching (Fig. 2b,c).

**RP2**. Adding, besides Shannon noise limitation, the threshold of perception, determined by the sensitivity of the receiver ($E_R$ in Fig. 2b,c).

**RP3**. Taking into account the rapidity of the receiver – the fact that the receiver can switch both much faster (the right part of the RS's existence region in Fig. 2c) and much slower than the signal (left side of Fig. 2c); in the second case the receiver integrates the signal.

**RP4**. Introduction of a threshold limitation (*H*) for integrated signals.

**RP5**. Complete elimination of infinities by introducing in addition to thresholds ($E_R$, *H*) the physical limit (*P*) of signal also. The intersection of the lines ($E_R$, and *P*) gives the point of greatest possible rapidity of the receiver $\delta t_M = E_R/P$ (Fig. 2b), and in the case of integration gives the point of greatest possible speed of the signal recording $\delta t_M = \sqrt{H/P}$ (point '3' in Fig. 2c or point '2' in Fig. 2b).

*1.2.1. Action and interference of real signals*

At all seeming simplicity of interference, two circumstances are often overlooked. First: when summing the envelopes of signals, the total power $((\pm y_1)+(\pm y_2))^2$ is not equal to $(y_1^2+y_2^2)$. This implies a non-trivial conclusion [2]: <u>free</u> (without interaction) interference is possible only for *two-component* signals. Electromagnetic signals are two-component; as far as it is known, gravitational ones are one-component and cannot freely pass through each other.

Second: interference is always associated with the algebraic addition of *radicals*. In electrical equipment the power radicals are voltage and current. Action is also summarized [3, 4] through radicals:

$$\sqrt{H_{1+2+\cdots}} = \sqrt{H_1} \pm \sqrt{H_2} \pm \ldots \qquad (1)$$

In the microcosm, the probabilities (squares of the Ψ-functions moduli) also interfere through their radicals.

*1.2.2. Finiteness of state changing rapidity*

The restriction of the transition rapidity by the point '2' in Fig. 2b, means the absence of jumps and breaks in the transition, i.e. smooth transition of the signal from one state to another. From this follows that the signal between the discrete threshold levels gradually passes through the fractional threshold values (see the set of horizontal lines in Fig. 2b and the hyperbolas in Fig. 2c). Of course, such passing is impossible to register; moreover, such a transition mechanism can only be considered as <u>virtual</u> (possible).

For the threshold hyperbola in the coordinates $E$-$\delta t$, the threshold ratio has a simple form:

$$H = E \cdot \delta t, \quad (2a)$$

which can also be written in the spectral representation:

$$H = E / \Delta\nu = E / \nu_H. \quad (2b)$$

This level of *indefinability* is determined by the *physics of perception* of the signal by a *slow* receiver.

*1.3. Algebraic logic of virtual transitions*

The logical variables '... C, B, A' in steady states have integer values of 0 or 1. However, it is possible to use [9, 10] the algebraic variables '... c, b, a' provided that they take values only 0 or 1. In this case, logical operations (inversions, conjunctions and disjunctions) are reduced to subtraction from unit, multiplication or addition respectively:

$$A \Rightarrow a, \quad \bar{A} \Rightarrow 1 - a, \quad \text{(the same for 'b'...)} \quad (3a \& 3b)$$

$$A \wedge B \Rightarrow a \cdot b. \quad (3c)$$

$$A \vee B \Rightarrow \{1 - [(1-a)\cdot(1-b)]\} = a + b - (a \cdot b). \quad (3d)$$

One circumstance is extremely interesting and important [4]: the multiplication and addition of the variables '... c, b, a' (3c, 3d) coincide in form with the expressions for the probability of the product of independent events and for the probability of the sum of *joint* events:

$$p(A \cdot B) = p(A) \cdot p(B), \quad (4a)$$

$$p(A+B) = p(A) + p(B) - [p(A) \cdot p(B)]. \quad (4b)$$

Consequently, the variables of *algebraic logic* can take the values of the probabilities between zero and unit: $0 \geq ... c, b, a \leq 1$. Continuous variables '... c, b, a' in each cell of the Karnaugh logic card [11] – Fig. 3b, form the sub-card of virtual transitions for this cell (Fig. 3e). At a large number of variables, such a card is more conveniently represented as a table (matrix) – Fig. 3c,f. Details can be found in [3, 4].

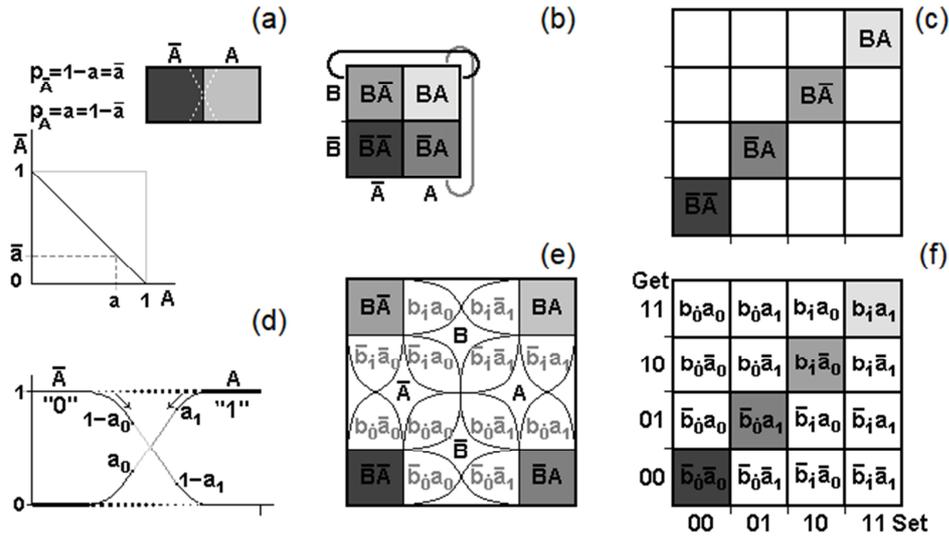

Fig. 3. (a) The classical transition by jump from A to $\bar{A}$ and back. (b) For two variables their logical combinations (here is the conjunction, its sign is omitted) are displayed by the Karnaugh card (it is considered that the card is a planar scan of closed surface, where each cell '... CBA' is adjacent to all others). (d) The transitions $A \leftrightarrows \bar{A}$ are smooth. (e) Then each combination has smooth transitions with all neighbors; as a result of which the number of cells in the card grows quadratically, and the card acquires a two-level structure (the upper one is the level of the variables '... CBA' and the lower one contains random transitive states. (c) In this case, it is more convenient to represent the Karnaugh card as a table (matrix). (f) The successors of the variables '... CBA' occupy the main diagonal in the table; the products $b_j \cdot a_i$ are the probabilities of random sub-threshold states, which summarized on columns arithmetically, as shown in the graph of fragment (a).

*1.4. The work objective*

It is believed that quantum mechanics (QM) is difficult to understand because its properties have no analogues on the scale in which we ourselves exist. Therefore, a large number of postulates have been introduced into QM basis. Comparative analysis, conducted in the introductory part of the paper, clearly shows that the problem is not in existing reality, but in its classical representation with infinite infinities. The aim of the work was to find out which of the properties of real signals that were introduced in the macrocosm may be revealed also in the basic concepts of QM. The following issues were considered:

a) What properties of the RS contribute, in contrast to the classics, their presence in the microcosm (this is partially discussed above);

b) What exactly express the presence of RS in the microcosm and their correspondence to the QM bases;

c) Whether the provisions of the RS theory can be useful in solving questions concerning QM.

An example in the last paragraph may be the fact that a real signal is capable of possessing action, but not every signal has it (its area may be equal to zero).

**2. Real signals in the microcosm**

If in QM the transition to the classics corresponds to $h \to 0$, then it means that a threshold on action in the *macro-scale* is impossible. However, the real signals with $H >> h$ exist; every doubter can look at any TV screen (sight is inert and the integration of 25 frames per second is obvious).

*2.1. Thermal noise as the boundary of the transition to QM*

Planck introduced the first postulate of QM for the disposal of classical physics from the "ultraviolet catastrophe":

$$E = h \cdot \nu. \qquad (5a)$$

The cutoff parameter of the frequency spectrum, in the distribution found by him, is the ratio $h\nu/kT$ in the (negative) power of the exponential function. The multiplication of Boltzmann constant $k$ by the absolute temperature $T$ in the denominator is the characteristic energy of the Nyquist classical thermal noise [12]. Everything told is well known and, nevertheless, raises the question: what relation the boundary of disappearance of classical *noise* $h\nu \approx kT$ has to quantum mechanics? Besides, in a formula (5a) precisely specified frequency '$\nu$' is present, i.e. the harmonic oscillation. Another example of using harmonic oscillation (although not in such a clear form) is the "pilot wave" postulated by de Broglie:

$$p = b \cdot \kappa / 2\pi, \qquad (5b)$$

where $b = h \cdot c$, and $\kappa$ is the wave number – an analogue in the spatial coordinate, of the circular frequency $\omega = 2\pi \cdot \nu$. "The shadow of the classics" is obviously present here, because harmonic has the beginning in minus-infinity only. For comparison: the same simply derived and not postulated formula (2b) for signals contains $\Delta\nu$ or its equivalent $\nu_H$ – the upper frequency of a *broadband* wave packet whose length is tied to energy.

In general, the situation is somewhat paradoxical: in classical physics, signals are present and widely used, and then they kind of disappear (together with classical noise). Although according to [3, 4], the dynamic signal is the first derivative of the *interaction potential* that continues to exist and affects the kinetic energy of the particle. Therefore, the problem here is not in the signals themselves, but in the fact that used abstract CMA tools cease to describe them. It can also be noted that the length of the dynamic signals in space is given by the speed of light ($c$), since with respect to the spatial variable the threshold constant $b = h \cdot c$.

## 3. Indefinability as a consequence of the perception threshold

The existence of a threshold and its numerical value are properties of the receiver of signals and not of their source. From this it follows that one cannot exclude the existence of sub-threshold signals, which cannot be perceived.

### 3.1. Evolution of the sub-threshold signal

Real signals allow us to estimate the dynamics of the signals transition from virtual states $a \leftrightarrows \bar{a}$ ($\bar{a}=1–a$) to the threshold level $A \leftrightarrows \bar{A}$. Suppose that there is a signal in the interval $t_0 \div t_1$, with the envelope given by the function $A(t)$. The fact that the signal in physical interactions is the radical of the interaction potential derivative can be expressed by the following formula:

$$A(r) = \pm\sqrt{dW/dr}; \qquad (6a)$$

in turn, as mentioned above:

$$A(t) = A(r)/c. \qquad (6b)$$

The signal energy is equal to the integral of the square of its envelope $A(t)$:

$$E_S = \int_{t_0}^{t_1} A^2(t) \cdot dt. \qquad (6c)$$

To find the action $H_S$ of the signal, its area

$$S = \int_{t_0}^{t_1} A(t) \cdot dt \qquad (6d)$$

should be squared: $H_S = S^2$.

According to [2÷4], the highest harmonics of the signal spectrum cannot be perceived due to the inertness of the recorder (receiver). The higher harmonics of the spectrum can be excluded at averaging of a signal by its particular integration within the interval $\tau$:

$$A_\tau(t) = (1/\tau) \cdot \int_{t-\tau}^{t} A(t) \cdot dt, \qquad (7a)$$

wherein $\tau$ may be either smaller or much larger than $\Delta t = t_1 - t_0$. In all cases, harmonics above the frequency $\nu = 1/\tau$ are eliminated. Total area of the integrated signal:

$$S_\tau = \int_{t_0}^{t_1+\tau} A_\tau(t) \cdot dt, \qquad (7b)$$

remains unchanged ($S_\tau = S$) because it is determined by the "zero harmonic" ($g_0$ – constant component) of the signal spectrum.

### 3.2. In what the threshold of perception is revealed

The indefinability relation (2b) is accurate, in the sense that repeated measurements under similar conditions will yield the same result. It may be compared with ray optics: although the image of the parts may be slightly blurred by defocusing, however, the resulting (photo)picture is static (does not vary from measurement to measurement – from photo to photo). If the optics allows, for example, when surveying the Earth from space, to distinguish the details of the surface, but the presence of an intermediate thermal (noise) air medium blurs them, then you can "pull out" the details by making a series of photographs and averaging the images. This effect of improving the accuracy of samples due to the mixing of sub-threshold noise (less than the division of the measuring scale) is widely used in nuclear electronics [13]. The *indefinability* relation gives stable results because it corresponds to the position of the reference points in Fig. 1e. But between the reference points there is an <u>uncertainty</u> which has a *random* character. The known "Heisenberg microscope" works in accordance with ray optics also. Its indeterminacy is related to the Fourier transform property and *does not depend on energy* [2], just as it has no relation to randomness.

As it was shown in [2], the *indefinability*, in the form of the relation (2b), is present in a sharp-cornered potential well. Besides, the parameters of the first orbit of the hydrogen atom are derived in [14] from (2b) at $H=h$. First of all, this indicates that the *indefinability* relation works really. Secondly, this means that the ratio obtained in the theory of signals corresponds to Bohr's postulate about stationary orbits.

*3.3. The reason for the appearance of the tunnel effect*

It is believed that the tunnel effect follows from the solution of the QM wave equation only, therefore it is unusual and inexplicable in the macroscales, i.e. with $H>h$. Below will be shown that this effect is a simple consequence of the inertness of (real) signal perception.

The signal, with the trapezoidal potential of interaction – Fig. 4a, has a quasi-rectangular form – Fig. 4b. At the very beginning of the signal ($t \approx t_0$), its energy and action will be small. Since the density of the action (in time) is energy, then with the passage of time ($t>t_0$) the action will grow quadratically ($A(t)$=const). At the end of the signal, its energy becomes constant, and the action will continue to grow, but linearly: $\delta H_S = E_S \cdot \delta t$. Therefore, with any threshold on action $H_R$ of the receiver, such a signal will be recorded (perceived) if its integration constant $\tau_R$ is sufficiently large: $E_S \cdot \tau_R \geq H_R$. If the interaction potential changes frequently, it creates a set of signals; they will be registered *jointly* – according to the addition rule of actions (1) in the integration zone.

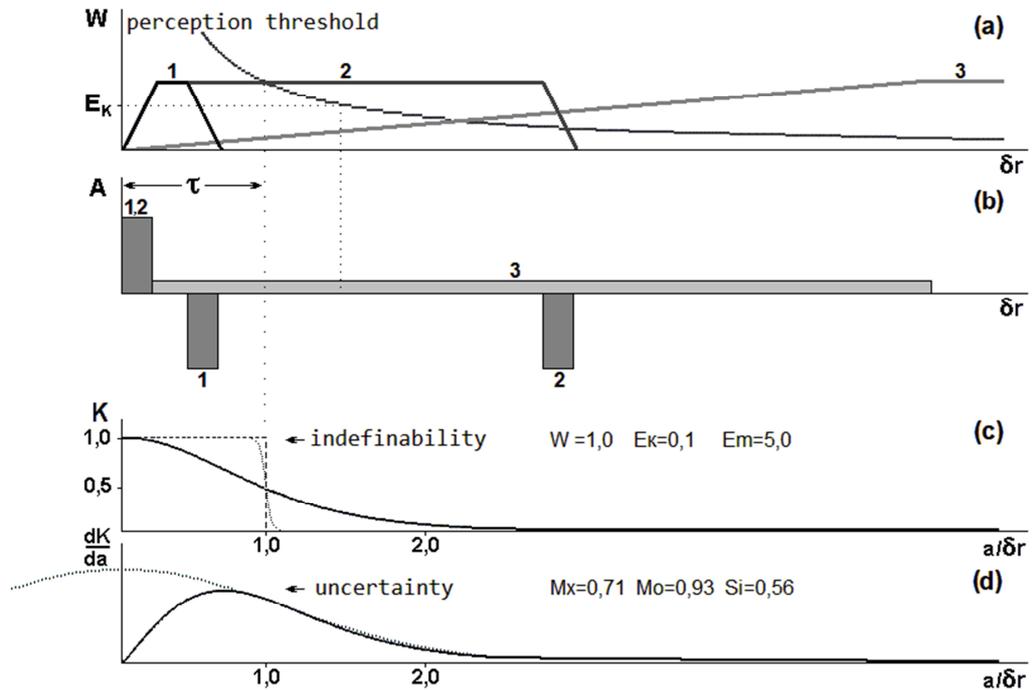

Fig. 4. (a) Three variants of potential barriers. (b) Signals are derivatives of their level differences. The first barrier is much narrower than the integration zone; therefore a couple of its differently polar signals (rise and recession) has a total action equal to zero. In the second barrier, which is wider than the integration zone, there is no tunneling practically, but there must be a delay in the reflection of the particle. The third barrier is quasi-classical one. (c) Stepped threshold of perception of the barrier is shown by dashed straight lines. (c)÷(d) It is blurred by near-threshold noise with a probability distribution – are shown by solid lines. (d) For comparison, a Gaussian symmetrical (bipolar) distribution of classical noise is shown in a thin dotted line.

If the integration constant $\tau_R$ is sufficiently small, then the action of $H_S$ cannot reach $H_R$; such a signal will be sub-threshold for the receiver. Its *virtual* states will not be perceived directly.

Figure 4a shows 3 potential barriers – small (*1*), medium (*2*) and large width (*3*). The first barrier is most interest (in Fig. 4a,b is marked with '1'), where the particle is a "slow receiver" of the signal. This means that the barrier is significantly narrower than the signal integration zone (up to the point of intersection of the horizontal line $E_к$ with the threshold hyperbola $H$). Therefore, the signals from its front and recession (Fig. 4b) are opposite in polarity and, in sum, neutralize each other. Signals from the second barrier are not summarized, but the particle must reach the threshold hyperbola before turn back. Finally, the third barrier is so sloping that it is close to the classical one.

The dashed rectangle in Fig. 4 shows the transition of a potential barrier from a transparent to an opaque state with increasing (from zero) the width of the barrier. As may be seen, this simple threshold mechanism for the perception of integrated signals works and explains the transparency of the *narrow* barrier. Therefore, there is no reason why this process cannot be realized in macro-scales. The macro receiver just has not enough time to detect a signal, although signal energy is quite large.

But in reality, not everything is so simple: the transition itself has rather stepped than smooth character – there is no virtuality noted in it (*indefinability* does not imply the presence of randomness). Intuitively it is easy to imagine that a virtual macro transition will look approximately as shown with the thin dotted curve in Fig. 4c.

## 4. Uncertainty as a consequence of noise – randomness

Thermal noise is destroyed under integration because it is bipolar (see the dashed line of the Gaussian distribution in Fig. 4d) therefore its area is zero and the action is zero respectively. In order noise is un-destroyable, it must be unipolar, i.e. its <u>m</u>athematical <u>e</u>xpectation (ME) should be greater than zero, and the distribution below ME should be power law, and not exponential.

Let us imagine that, in addition to the signal in Fig. 4a, there is unipolar noise. It is clear that it will be integrated, together with sub-threshold signals, repeating from event to event. Such noise will be for them a kind of "random elevator" to the threshold level. If, during the integration, the reached peak value of the signal action:

$$\hat{h}_\tau = (\int_{t_0}^{t} A_\tau(t) \cdot dt)^2, \tag{8a}$$

is less than $h$, it is clear that registration probability of "elevated" signal will be the higher the closer $\hat{h}_\tau$ to $h$. This can be represented in the following form:

$$p = \hat{h}_\tau / h. \tag{8b}$$

*4.1. Noise on action*

As shown in Fig. 2c, the <u>*fast*</u> signals (to the left of p.2) have a threshold on action. The second relation after (5a), similar in form to (2a), is the Heisenberg <u>uncertainty</u> relation for *fast* signals:

$$<\delta E^2> \cdot <\delta t^2> \geq (h/2)^2 \tag{9a}$$

The minimal action in the ratio (2b) is $h$, and in (9a) it is equal to $h/2$. The opinion that the formula (2b) is "approximate" and (9a) is "exact" formula of uncertainty relation has already passed into textbooks [15, 16]. Although in the already mentioned book [14] on the contrary (2b) is called the uncertainty relation and is claimed that it determines the radius of the first orbit of the hydrogen atom. The radii of the orbits obey Bohr's postulate – these are not approximate, but exact (classically determined) values. In this case what may be common between the exact "$h$-relation" (2b) and statistical "($h/2$)-relation" (9a), reflecting random deviations from the mean? Indeed, these are different in physical sense, although there is a relationship between them: the exact ratio (2b) is given by the threshold of perception (indefinability in Fig. 4c), and noise (uncertainty in Fig. 4d) is blurring it, see 4.1.1.

It is known that the density of the normal distribution of two random uncorrelated quantities is equal to the product of the densities of their distribution, therefore (9a) can be written jointly for $<E>$ and $<t>$ as follows:

$$<H> = h/2. \tag{9b}$$

Since the blur interval in (9a) in time is $\sim 2<\delta t>$, we can assume that $\Delta \nu = 1/2<\delta t>$, then from the obvious expression

$$\delta P = \delta E / \delta t = \delta H / \delta t^2 \tag{10a}$$

we obtain the average power of action-noise:

$$<P_\text{H}> = (h/2) \cdot (2\Delta \nu)^2 = 2h \cdot \Delta \nu^2. \tag{10b}$$

For comparison, the average power of classical (Nyquist) energy-noise [12]:

$$<P_E> = 4kT \cdot \Delta\nu. \quad (10c)$$

At the point '2' in Fig. 2, the following equality must be fulfilled: $<P_E>=<P_H>$; then from (10b) and (10c) follows that

$$h \cdot \nu_H / kT \cong 2. \quad (\nu_H \equiv \Delta\nu) \quad (10d)$$

This transition point is rather conditional because the transition is smooth really, but not stepped. At $T=300°$ K, the width of the (classical) noise spectrum is $\Delta\nu=2{,}6 \cdot 10^{13}$ Hz, and its average power $<P_E>=<P_H>=2{,}1 \cdot 10^{-7}$ W. Unipolar noise by action does not disappear, and increases with decreasing switching time of the signal $\delta t=1/\Delta\nu$, the latter is essentially limited by the point '3' in Fig. 2c.

*4.1.1. The perception threshold blurring by noise*

Let us now return to the potential barrier. The solid line in Fig. 4c is the probability function of the transition from the transparent state to the opaque, derived in [16] for a stepped barrier (i.e. the signals in Fig. 4b are represented by delta-functions):

$$D = 4/\{[(\kappa/\varkappa + \varkappa/\kappa)^2 \cdot sh^2(a \cdot \varkappa)] + 4\}. \quad (11a)$$

Here 'a' is the width of the barrier; '$\kappa$' and '$\varkappa$' are the wave numbers giving the square of the threshold length:

$$l_1^2 = (2\pi/\kappa)^2 = 4\pi^2/(2m \cdot E/\hbar^2), \quad \text{at } W=0; \quad (11b)$$
$$l_2^2 = (2\pi/\varkappa)^2 = 4\pi^2/(2m \cdot E((W/E)-1)/\hbar^2), \quad \text{at } W>0; \quad (11c)$$

and the hyperbolic sine: $sh(a \cdot x) = (\exp(a \cdot x) - \exp(-a \cdot x))/2$.

It's clear that the previously obtained stepped indefinability threshold is indeed blurred by uncertainty (by noise). Fig. 4d shows the derivative of the distribution function, i.e. probability density distribution:

$$dD/da = -(D^2/4) \cdot [(\kappa/\varkappa + \varkappa/\kappa)^2 \cdot (\varkappa/2) \cdot (\exp^2(a \cdot \varkappa) - \exp^2(-a \cdot \varkappa))]. \quad (11d)$$

The delta-function – the derivative of the stepped threshold in Fig. 4c, is blurred here. As was mentioned above, using the indefinability relation, one can find the reference points of motion "with increment on $h$". Using the uncertainty relation, one can find a noise blur between these points "with a statistical deviation of $h/2$".

Let's see now what happens to logical cards and state tables (Fig. 4d, e), if the signals are fully integrated.

*4.2. The joint states of the fast signals*

The particular integration of the signals (7a) leads to:
a) Excluding the highest harmonics only;
b) The appearance of virtual transitions (dotted line in Fig. 4c);
c) A small noise blur in the *columns* of the state table – Fig. 4e.

Full integration – up to the lowest harmonics ($\tau >> \delta t_R$), leads to the appearance of a blur not only inside the columns, but also between them. As a result, joint virtual states appear [4], i.e. superposition states of QM, for example, in qubits [17, 18].

The superposition of one qubit states, in terms of algebraic logic, can be written as $\overline{a}|0> + a|1>$, where $a$ and $\overline{a}$ in Fig. 5a represent the mnemonic designation of numbers whose sum of squares is equal to 1. For 2 qubits there is a superposition of four base values: $\overline{b}\,\overline{a}|00> + \overline{b}\,a|01> + b\,\overline{a}|10> + b\,a|11>$, shown in the table in Fig. 5c. In the previous sentence the numbers $b_j a_i$ are the identifiers of the probability radicals, and not the multiplications $b_j \cdot a_i$, as in the table in Fig. 3f. This kind of notation is used, instead of the usual index of a number, in order to show the similarity of the structure and the "hereditary relation" of the table in Fig. 5c with the table in Fig. 3f. In the table in Fig. 3f the probabilities were given; the indication with radicals did not make much sense there, since interference (in rows) is practically absent. An example of the interference of probability radicals [19] is given in Fig. 5d.

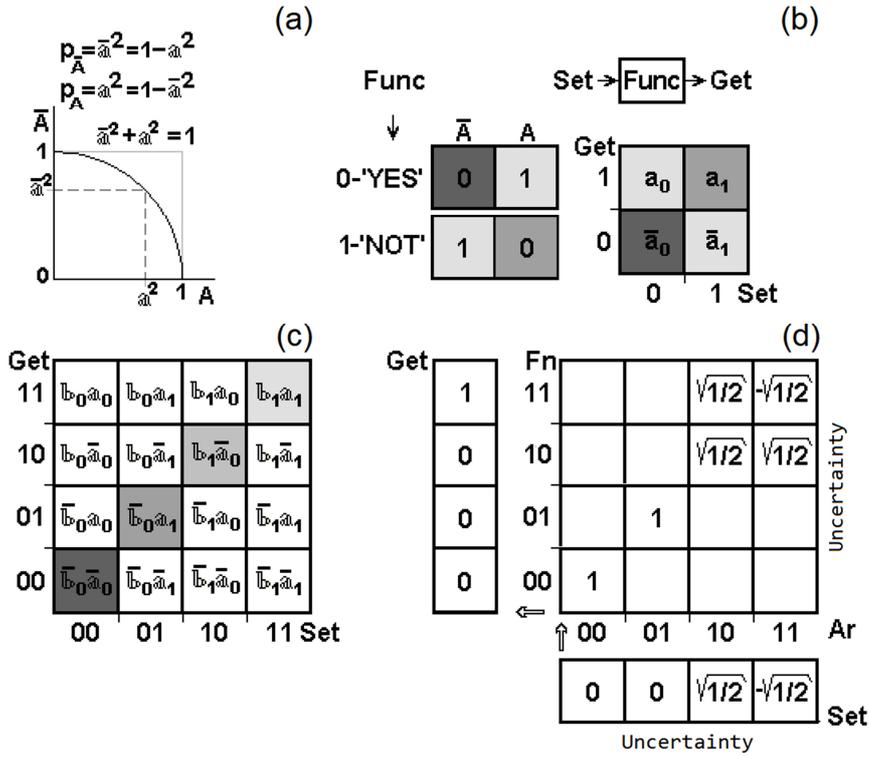

Fig. 5. (a) Geometrical summation of probability radicals. (b) At the left – the 'YES' and 'NO' functions of one classical logical variable; at the right – a table for functions of one algebraic logical variable (potential barrier, qubit). (c) The table of joint states of two variables, the numbers $b_j a_i$ are the probability radicals. 'Ar' is the argument; 'Fn' is the function of conversion. (d) Get-states are the result of interference of probability radicals in rows of the table.

*4.3. The relationship between the probability of a virtual state and a sub-threshold action*

Interference means that there is an algebraic summation of the probability radicals:
$$\sqrt{p_1} \pm \sqrt{p_2} \pm \ldots, = \sqrt{p_{1+2+\ldots}}, \tag{12a}$$
where $p_i$ are the probabilities of particular states, and $p_{1+2+\ldots}$ is the total probability for one row of the table in Fig. 5d.

Let us compare formula (12a) with formula (1) for the addition of action radicals [4]; last one is normalized to the radical of the threshold action $h$:
$$\sqrt{\hbar_{1+2+\ldots}} / \sqrt{h} = (\sqrt{\hbar_1} \pm \sqrt{\hbar_2} \pm \ldots) / \sqrt{h}, \tag{12b}$$
here $\hbar_{1+2+\ldots}$ is the total action. From a comparison of (12a) and (12b) it can be concluded that:
$$\pm\sqrt{\hbar_i/h} \equiv \pm\sqrt{p_i}. \tag{12c}$$
Raising both sides of (12c) into a square, we obtain the same relation as in formula (8b) for probability and action. Note that the probability radical $\sqrt{p_i}$ is a real number, i.e. module of the complex "probability amplitude". Although it is asserted that the $\Psi$-wave in QM is "not similar to a real wave" and "no reality can be associated with this wave" [20] formula (12c) reflects precisely the physical sense of this "wave".

Even from the brief analysis given above, it is clear that, in contrast to the classical ones, the properties of real signals are rather well combined with the beginnings of QM, because they have a perception threshold in the zone of signal integration (from p.2 to p.3). In addition, formulas (8b) and (12c) show that sub-threshold states can be registered at the presence of randomness, but require to collect statistics.

Notice that the sub-threshold noise states are not exist simultaneously; the possibilities only are present for occupying them with one or another probability. Note also that reasoning within the framework of classical integer logic 0/1 are suitable only for quantities that have practically no noise. If noise is present, then is necessary, including in macro-scales, to use the algebraic logic of probabilities 0≤p≤1. Its means that "logically" talk about matters is impossible if they are unknown and *unpredictable* to us.

Since ignorance creates randomness, it is necessary to take into account all possible variants of events.

## 5. Threshold values in the wave equations of QM

Threshold variables of real signals [2] for the coordinate and time have the form:
$$\delta r = \bar{h}/E_p, \qquad \delta t = \hbar/E, \qquad (13a \& 13b)$$
here $\bar{h}=\hbar \cdot c$, $E_p=p \cdot c$, where $p$ is the momentum of the particle, and $E$ is its total energy. As is known, $(E–W)$ is the algebraic sum of $E_m$ and $E_к$:
$$E - W = E_m + E_к, \qquad (14a)$$
and the "geometric" sum of $E_m$ and $E_p$. Here $W$ is the potential energy of the particle, $E_к$ is its kinetic energy, and $E_m=m \cdot c^2$. Let represent (14, a) in the following form:
$$E - E_m = W + E_к = E_t, \qquad (14,б)$$
and assume that $E_m$=const; it means that the internal energy of the particle does not change. Then, in a closed ($E$=const) classical system $W+E_к$=const, because a compensating change of $E_к$ follows immediately after a change in the potential W: $\delta W+\delta E_к=0$. It is known [20] that in micro-scale the rapid W change (fast signals) is disturbed this balance randomly (by noise – within the indefinability relation).

Since all subthreshold evolutions occur only with the sum $W+E_к$, the canonical wave equation for the threshold variables (13a, b) can be written [2] in the form:
$$(\bar{h}^2/E_p^2) \cdot (\partial^2 \Psi / \partial r^2) - (\hbar^2/E_t^2) \cdot (\partial^2 \Psi / \partial t^2) = 0. \qquad (15,a)$$
The wave in QM
$$\Psi(r, t) = \exp[i \cdot ((E_p \cdot r/\bar{h}) - (E \cdot t/\hbar))]. \qquad (15b)$$
is a solution for stationary wave equations. It was shown in [2] that from (15a, b) it is easy to obtain both the nonrelativistic Schrödinger equation and the relativistic Klein-Gordon equation ($E_p$, $E_t$ and $E$ must be represented in a nonrelativistic or relativistic forms, respectively).

As was already mentioned the indefinability takes place in a potential box with "rigid" walls [2]. The oscillations of the Ψ-"wave" in it are harmonic due to self-interference of counter-propagating harmonics moving with the *phase* velocity $v_н$.

Note that in order to obtain a signal representation of the "gap" through which the "receiver"-particle passes, it is only necessary to eliminate the higher harmonics in the spectrum of its image (Fig. 1c and Fig. 1d for a narrow and wide gap, respectively). The solution of this task in QM is more complicated [21].

## 6. Limit restrictions of real signals

It can be seen from Fig. 2b,c, the threshold limitation is clearly not enough to describe by signals the real phenomena, even with the addition of the 2-3 (QM) region. It is necessary to take into account the limit restrictions for the full description. The points '3' and '4' (Fig. 3b,c) will be absent without the existence of a limit and the rapidity of signal switching becomes infinite. In addition, as is evident from Fig. 1f, at the density limit the energy begins to stretch "into a string".

If the threshold limitation at energy increasing is associated with the expansion of the signal spectrum, then the increase of energy above the point '3' in Fig. 2c ($E_M$) is associated with signal slowdown (more precisely, with its expansion): in Fig. 1c $\Delta t=2 \cdot \delta t_M$, and in Fig. 1f $\Delta t=16 \cdot \delta t_M$. The threshold and limit lines intersect in p.3, where the power limit
$$P_M = h / \delta t_M^2 = E_M / \delta t_M. \qquad (16a)$$

With the energy $E$ increases (the upper part of Fig. 2c), the power limit remains constant: $P$=const; then the duration of the limiting (on power) signals grows linearly:
$$\Delta t = E / P, \quad (16b)$$
and their action is quadratic:
$$H = E^2 / P. \quad (16c)$$
It is significant that for gravitational interactions the point '3' of the intersection of the threshold and limit lines ($E_M$, $\delta t_M$) is the Planck point. He obtained this point on the basis of a comparison of the dimensions of the fundamental constants. In [2] this point was *derived* as the limit of the signal switching rapidity (creating an information limit) not only in gravitational interactions. Note that at the level of the limit, only the length of the signal carries information. The values of $E_M$ and $\delta t_M$ for different types of interactions are given in [2÷4].

## 6.1. Dispersion as the aging of real signals

The restriction on the switching rapidity of the signal – $\delta t_M$, corresponds to the restriction in the spectrum $\nu_M \approx 1/2\delta t_M$ and inevitably leads to the dispersion of single signals during motion. In the first approximation, the phase shift of the higher-pass filter - Fig. 6a, has the form [3]:
$$\phi(\nu) \cong (\nu / \nu_M) + (\nu / 2\nu_M)^3. \quad (17a)$$
The first term provides a delay of the signal, and the second adds nonlinear phase distortions to it, that increasing with the increase of the traversed path $r$:
$$\delta\phi = (\nu / 2\nu_M)^3 \cdot r / (c / 4\pi\nu_M). \quad (17b)$$
In Fig. 6b this is shown for the sampling function (*1*) as a signal with three increasing distances (*2÷4*). The height of the central peak of the curve (*2*) is less than in original peak (*1*) and corresponds to a 10% decrease in its energy.

In Fig. 6c the curve (2h) is the sum of the lagged behind higher harmonics (10% of energy), it is a narrowband wave packet with the action equal to zero. This means that it becomes "invisible" (dark) and cannot be registered.

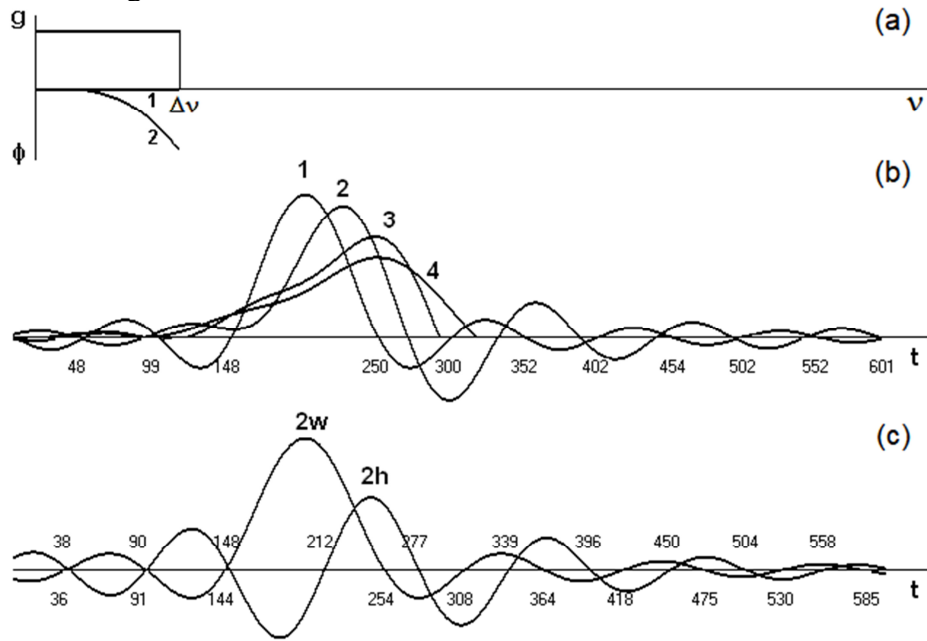

Fig.6, (a) The flat spectrum (width $\Delta\nu = \nu_M$) of the sample function and an example of the phase shift curve of its harmonics. (b) The envelope shape changing of the original signal (1) with an increasing nonlinear delay of its harmonics (2÷4, for the curves 3÷4 the central peaks are shown only). (c) The allocation of a narrowband packet of higher, significantly lagging harmonics of the signal (2h) and the remainder of the broadband signal (2w). The numbers indicate the times of zero line crossing by the signal (1) - (b) and by both parts of the signal (2) - (c).

The energy of the main part (broadband) of the packet decreases, but remains registerable. If assume that in the visible range the wavelength of the photon is ~$6 \cdot 10^{-7}$ m, then according to (17b), a 10-percent drop in the effective energy (aging) of the photon will occur after a distance of L=$0,4 \cdot 10^{25}$ m has passed.

## 7. Conclusion

In a classical signal (more precisely, in a classical mathematics) a property is declared: the deviation of each point can be from infinitesimal to infinitely large. Each time, other signals different from the original (f(x)=0) are obtained, *devoid of energy*, but *having power*. The information is transmitted by each of them, since point states (for mathematicians) are *distinguishable*.

The key means of resolving classical contradictions in signals is the elimination of infinitesimal and large values in signals (Planck's '3' point in Fig. 2c), as well as in their derivatives (elimination of jumps).

Although the QM perfectly describes the quantitative aspect of the phenomena, the answer to the question "why this happens" may remain in the shadow. This is clearly visible on the large number of its postulates. The motivation for this work was to identify the relationship between the formulas derived in [1÷4] for real signals and the basic postulated relations in QM. On this side, Planck's postulate *E=h·ν*, Bohr's postulates on stable states and frequency conditions, de Broglie's hypothesis about the "wave-pilot" of the particle, the postulated Schrodinger wave equation, the measurement postulate were touched.

Now we can answer the three questions that were defining the work purpose in the introduction. For the beginning, we summarize the prerequisites provide the properties of actual signals:

(a) The partial integration of the signal removes the highest harmonics of its spectrum; full integration removes the entire spectrum of the signal except for the constant component $\pm g_0$, the last one is equivalent of a signal area and has polarity.

(b) The signal area square is action, therefore areas of the signals (as action radicals) can interfere; the conservation of energy of the "free" interference in a closed system additionally requires the presence of two orthogonal components.

(c) The algebraic logic of the virtual states of the RS coincides with the expressions for the probability of the product of independent events and with probability of the sum of joint events; for weakly blurred signals, it goes to the classic integer logic (0/1).

1. *What properties of RS ensure their existence in the microcosm*

1.1. The main features of real signals are as follows.

1.1.1. Real signals are obtained from the classical ones by eliminating the infinity of the spectral (from below) and linear (from above) densities of their energy.

1.1.2. Any signal receivers (including atoms and particles) are inert and have a fundamental perception threshold
- for *slow* signals, this is the threshold on energy, the signal is only partially integrated;
- for *fast* signals this is the threshold on action, the signal itself is fully integrated.

The change in the state of any, even a fast receiver, cannot be instantaneous.

1.2. The spectrum of the RS is limited and has a finite number of reference points carrying information. The higher harmonics of the spectrum, although small in amplitude, give the same contribution to the number of reference points on the envelope of the signal, as the harmonics of the main part of the spectrum.

1.3. In micro scales in the boundary region of frequencies $\nu_H \approx 2kT/h$, the threshold properties of real signals correspond to the transition from classical laws to quantum ones.

2. *In what the presence of RS is expressed in a microcosm*

2.1. Indefinability of the highest frequencies of the signal spectrum.
2.2. The presence of noise on action. There are only *opportunities* in the noise to occupy any state at the same time.
2.3. The existence of "algebraic" non-integer logic.
2.4. Aging of the signals and formation of "invisible" ($\hbar \ll h$) narrow-band energy packets.
2.5. The presence of restrictions on the power $P_M$, leading to a "stretching" of the signal energy.

3. *Can the provisions of RS-theory be useful in matters relating to QM foundations?*
3.1. In RS-theory is rather simply deriving the ratio $E=h\cdot\nu_H$, which was postulated by Planck.
3.2. The condition $2kT=h\cdot\nu_H$, which defines the boundary of classical noise, is obtained simply enough.
3.3. It is shown that the transition from $E_N=4kT$ to $E_N=h\cdot 2\nu_H$ is related to the integration of classical signals and that the QM phenomena are located in the region of fully integrated signals.
3.4. The transition from the postulated harmonics (Planck - $E=h\cdot\nu$, de Broglie - $p=b\cdot\kappa/2\pi$) to the broadband wave packet ($\Delta\nu$, $\Delta r$), where the highest harmonic is characteristic, and the zero-"harmonic" ensures the presence of an action.
3.5. The noise on action, that blurring the indefinability, is distinguished and its relationship with the statistical uncertainty relation is shown.
3.6. The representations of RS-theory can complement highly developed QM instruments, giving a more visual and less formal interpretation of the physical meaning of the described processes. For example, in the RS-theory *indefinability* is introduced as a threshold of perception.
3.7. Algebraic logic is introduced for virtual transitions and joint (superposition) states. The integer logic (0/1) corresponds only to the states that have not a blur.
3.8. Sub-threshold variables are associated with sub-threshold action and with radicals of probability of joint (superposition) states: $\pm\sqrt{\hbar_i/h} \equiv \pm\sqrt{p_i}$.
3.9. The point '2' of the transition from the threshold of perception on energy to the threshold of perception on action determines the maximum speed of information transfer.
3.10. Any narrowband ($\delta\nu$) wave packets (wavelets) that do not have a "zero-harmonic" $g_0$ (constant component) have no action, since their area is equal to zero.
3.11. Physically such packets may be formed during real signals dispersion; their registration by means with the threshold on action equal to $h$ is very problematic.
3.12. The description of the full range of physical phenomena requires, in addition to the threshold limitations of the QM, the fundamental limitations of maximum density of energy also.

In general, using RS opens partial similarity in micro and macro effects. In classics, the two-level state table is filled only over the (main) diagonal. In partially integrated real signals, each of the ground states can be blurred by noise in its column; this creates the lower level in the table. With full integration of real signals, all states along the horizontal of the state table become joint (superposition) states and can interfere with each other through the radicals of action. The property of superposition, as a consequence of the integration, and the presence of noise on action, are exceptional properties of states in QM (the first corresponds to the well-known statement of Dirac). Since the noise blur of the table begins already in the macro states of real signals, the transition from "macro" to "micro" becomes not as sharp as in the "classics" and more understandable.

Addition to the classical theory the real concepts of the *inert* threshold of perception and the limit of the signal existence transforms the classics into a general theory of signals (OTS) devoid of infinity.